\documentclass[twocolumn,showpacs,preprintnumbers,amsmath,amssymb]{revtex4}
\usepackage{graphicx}% Include figure files
\usepackage{dcolumn}% Align table columns on decimal point
\usepackage{bm}% bold math

\begin{document}
\input epsf

\title{Dipolar corrections to the static magnetic susceptibility of condensed $^3$He}
\author{N. Gov}
\address{Department of Materials and Interfaces,
The Weizmann Institute of Science,\\
P.O.B. 26, Rehovot, Israel 76100}

\begin{abstract}

We examine the consequences of a recent model describing
correlated zero-point polarization of the electronic cloud in
solid $^3$He. This polarization arises from the highly anisotropic
and correlated dynamic mixing of the $s$ and $p$ electronic levels
($\sim1$\%). The magnetic polarization introduces a small
paramagnetic correction, of $1-0.1$\%, to the static
susceptibility of condensed $^3$He. This correction could explain
recent measurements in liquid $^3$He.

\end{abstract}

\maketitle

Recent experiments \cite{gould} have called attention to the
problem of the diamagnetism of condensed phases (liquid and
solids) of $^3$He. Theoretical calculations \cite{bruch} have not
been able to reproduce the relatively large ($\sim 1\%$) observed
paramagnetic deviation from the expected value of the electronic
diamagnetic response. In the ground-state of the free $^3$He atom
$\left| 0 \right\rangle$ there is no electronic magnetic moment,
as the filled $s$-shell has the two electrons in a singlet state.
The static magnetic susceptibility of an insulating material such
as $^3$He is given by \cite{ashcroft}
\begin{eqnarray}
\chi&=& -\frac{N}{V}\frac{e^2}{4mc^2} \left\langle 0 \right| \sum_{i} \left( x_{i}^2+y_{i}^2 \right) \left| 0 \right\rangle \nonumber \\
& &+2\mu_{B}^2 \frac{N}{V}\sum_{n} \frac{ \left| \left\langle 0 \right| \left( {\bf L_{z}}+g_{0}{\bf S_{z}} \right) \left| n \right\rangle \right| ^2}{E_{n}-E_{0}} \label{suscept}
\end{eqnarray}
where $g_{0}\simeq 2$, $\mu_{B}=e\hbar/(2mc)$ ($m$ is the electron
mass) and the applied magnetic field is taken to be in the
$z$-direction, so that only the perpendicular ($x,y$) projection
of the electronic cloud contributes to the first, diamagnetic,
term, and only $L_{z},S_{z}$ operators appear in the second,
paramagnetic term.

The dynamic corrections to the susceptibility where recently
calculated \cite{bruch}, and found a $\sim 0.1\%$ diamagnetic
enhancement, with much smaller paramagnetic terms. This
calculation was carried out under the assumption of spherical
symmetry of the surrounding atomic potential. In our recent work
on low density solid He, we showed that in a highly anisotropic
potential the local zero-point atomic motion is correlated with
electric dipoles arising from the corrections to the
Born-Oppenheimer approximation \cite{afm}. This model allows us to
resolve long-standing problems regarding the experimental data in
bcc $^3$He and $^4$He \cite{afm,niremil}. In addition we have
predicted the appearance of a new excitations mode, which was
recently observed by neutron scattering \cite{emiltuvy}. Following
this experimental evidence supporting our model, it seems
worthwhile to explore the consequences for the static magnetic
susceptibility of He. We show here that our model gives a
paramagnetic correction, which is in rough agreement with the
observed value of $\sim1 \%$.

We shall now give briefly the main properties of the ground-state
dynamic mixing of the $s$ and $p$-shells in solid He (see more
details in \cite{afm}). In the low density (specific volume $V<21$
cm$^3$/mole) bcc phase of He, the potential an atom feels due to
the Van-der Waals interactions with its neighbors, has a
double-well structure \cite{niremil,glyde}, along directions
pointing at the next-nearest neighbors. This is due to the average
inter-atomic separation ($\sim 3.5$\AA) being larger than the
minimum of the potential well ($\sim 2.7$\AA). Since the potential
well is highly anharmonic only along these particular directions,
the atoms will make highly directional, large amplitude zero-point
oscillations between these double-wells, with typical energy
\cite{niremil} $E_0\sim 1-10$K. We treat this anharmonicity as
decoupled from the higher energy modes, which are well described
by the usual harmonic approximation \cite{niremil,afm}.

We next couple the anaharmonic atomic motion with the atomic
polarizations due to the correction to the Born-Oppenheimer
approximation \cite{afm}. The $s-p$ level mixing is proportional
to $\lambda\propto\sqrt{m/M}\sim1\%$ (where $M$ is the nuclear
mass). Such a mixing results in an electric polarization of the
electronic cloud, i.e. a zero-point electric dipole moment:
$\mu\sim 2\lambda e \left\langle x\right\rangle$. When these
dipoles (and the atomic motion) on neighboring atoms are
uncorrelated, they give a negligible ($\sim0.1\%$) correction to
the usual Van-der Waals interactions.

If the atomic oscillations, and the corresponding electric
dipoles, are phase correlated, the crystal can lower its ground
state energy by choosing the correct relative phase for the
oscillation of adjacent atoms \cite{niremil}. For the bcc lattice
there is long-range order of these relative phases, and the
ground-state of the crystal may be described as a global state of
quantum resonance between two degenerate configurations which
minimize the instantaneous dipolar interaction energy
\cite{niremil}. In our picture, the zero-point energy is
resonantly transferred back and forth between the atomic motion
and the electronic energy in a coherent way throughout the
lattice, with the resonance frequency $E_0$.

In the hcp and liquid phases we do not expect there to be
long-range order in the dipole field, for the following reasons.
In the hcp solid the triangular symmetry frustrates long-range
ordering of the dipoles \cite{niremil}. In the liquid, the
direction of the instantaneous dipole on each atom is random, so
that local correlations should decay with distance even faster
than in the hcp solid. Additionally, the double-wells at different
sites are randomly spread over a range of energies $E_0$, causing
the oscillating dipoles to drift out of synchrony. Still, in both
these phases we expect local correlations between the dipoles to
exist. In order to describe single-atom properties, such as the
static susceptibility $\chi$ (Eq.\ref{suscept}), we will proceed
essentially as for the bcc phase.

The first contribution of the $p$-level mixing to the
susceptibility is in the diamagnetic (first) term of
Eq.\ref{suscept}. Due to the increase in the spatial extent of the
electronic cloud the diamagnetism is enhanced by
\begin{equation}
\chi_{dia}=\chi_{0}\left(1+2\lambda ^2 \frac{\left\langle p
\right| x^2+y^2 \left| p \right\rangle}{\left\langle s \right|
x^2+y^2 \left| s \right\rangle} \right) \label{dia1}
\end{equation}
where $\chi_{0}$ is the usual filled $s$-shell diamagnetism. Since
the ratio of the mean-square spread of the $p$ and $s$ shells is
$\sim7$, the correction (\ref{dia1}) is $\sim 0.1$\% (using
$\lambda \sim 10^{-2}$, as applicable for the lowest density solid
\cite{afm} and liquid). A correction of this magnitude was also
derived in \cite{bruch}.

We now describe an additional correction to the magnetic
susceptibility due to the dynamic and highly anisotropic mixing of
the electronic levels. The electric polarization of the electronic
cloud, has a magnetic polarization associated with it \cite{afm},
due to the following argument: The mixed $p$-state of lowest
energy is that with the two electrons in the $S=1$ state, due to
electrostatic repulsion, of magnitude $\sim 0.25$eV. Additionally
the spins of the two electrons will tend to align with the axis of
the $p$-shell, due to magnetic dipolar interaction between them.
The electronic spins are oriented with the oscillating $p$-states
of the atom, with effective energy $\left\langle E_{es}
\right\rangle \sim 1.4\cdot(1-F(V))$K, where the factor $1-F(V)$
takes into account the pressure-induced overlap of the different
electronic spin orientations \cite{afm}. This overlap reduces the
effective energy gap and electronic spin polarization:
$\left\langle {\bf \mu}_{e}(V)\right\rangle \simeq \left(
1-F(V)\right) \lambda (V) \mu_{B}$, where $\lambda(V)$ is the
volume dependent mixing coefficient. Since the overlap function
$F(V)$ is a very steep function of the specific volume $V$
\cite{afm}, and is $F(V)\sim0$ at the lowest density, we will take
it to be zero in the following calculation. Note that the
zero-point mixing of the $p$-shell with $S=1$ has zero static
expectation value of the magnetic and electric dipole moment.

The paramagnetic (second) term of Eq.(\ref{suscept}) is now
non-zero. Let us look at $p$-states orthogonal to the applied
magnetic field, i.e. in the $x,y$-directions. The mixed $p$-shell
of lowest energy is in the S=1 state \cite{afm}, with the spins
aligned along the axis of the $p$-state (Fig.1), i.e. $\left|
\psi_{0} \right\rangle \sim\left|s,s;S=0\right\rangle +\lambda
\left|s,p;S=1\right\rangle$, where $s,p$ stand for the orbital
angular momenta of the 2 electrons, and $S=0,1$ is the spin
angular momentum. This describes a mixing of the $p$-orbital along
the $x$-axis (for example), with the $s$-orbital, in a spin state
$S_{x}=\pm 1$. An excited state, with respect to $\left| \psi_{0}
\right\rangle$, is given by: $\left| \psi_{ex}\right\rangle
\sim\left|s,s;S=0\right\rangle +\lambda
\left|s,p;S=0\right\rangle$, where the $p$-orbital is mixed with
the $s$-orbital, in a spin state $S_{x}=0$. The $S_{z}$ operator
in the second term of Eq.(\ref{suscept}) connects between these
two levels for mixing in the $xy$-plane, so we get a term of the
form
\begin{eqnarray}
\chi_{para}&=&2\mu_{B}^2 \frac{N}{V}  \frac{ \left| \left\langle \psi_{0} \right| g_{0}{\bf S_{z}} \left|\psi_{ex} \right\rangle \right| ^2}{E_{S_{x}=0}-E_{S_{x}=1}} \nonumber \\
&=&2\mu_{B}^2 \lambda ^4 (1-F)^2 \frac{N}{V} \frac{ \left|
\left\langle S_{x}=1 \right| g_{0}{\bf S_{z}} \left|S_{x}=0
\right\rangle \right| ^2} {E_{es}} \label{dia2}
\end{eqnarray}
where $E_{es}=E_{S_{x}=0}-E_{S_{x}=1}\sim 1.4$K is the effective
spin-orbit interaction energy difference between the spin aligned
and anti-aligned states \cite{afm}.

Summing over a uniform distribution of the directions of the local
oscillating spins relative to the external magnetic field, the
value of this correction turns out to be
$\chi_{para}\simeq\chi_{0}\cdot10^5 \lambda^4\sim 1-0.1$\% (again
using $\lambda \sim 10^{-2}$). This is of the order of the
paramagnetic correction that was recently measured \cite{gould} in
liquid $^3$He.

To summarize, using our postulated dynamic and anisotropic $s-p$
mixing in bcc He, we calculate a paramagnetic correction to the
magnetic susceptibility. We argue that since the local atomic
neighborhood in the liquid is not very different from the low
density bcc solid, this paramagnetic correction should appear also
in the cold liquid \cite{cold}. The small size of the paramagnetic
correction, combined with the experimental uncertainties and
uncertainties in determining $\lambda$, make this effect hard to
calculate or observe with accuracy \cite{gould,bozler}.

\begin{acknowledgments}
I thank Gordon Baym for useful discussions. This work was
supported by the Fulbright Foreign Scholarship grant, NSF grant
no. DMR-99-86199 and NSF grant no. PHY-98-00978 while in the
University of Illinois at Urbana-Champaign.
\end{acknowledgments}

\begin{figure}
\centerline{\ \epsfysize 4cm \epsfbox{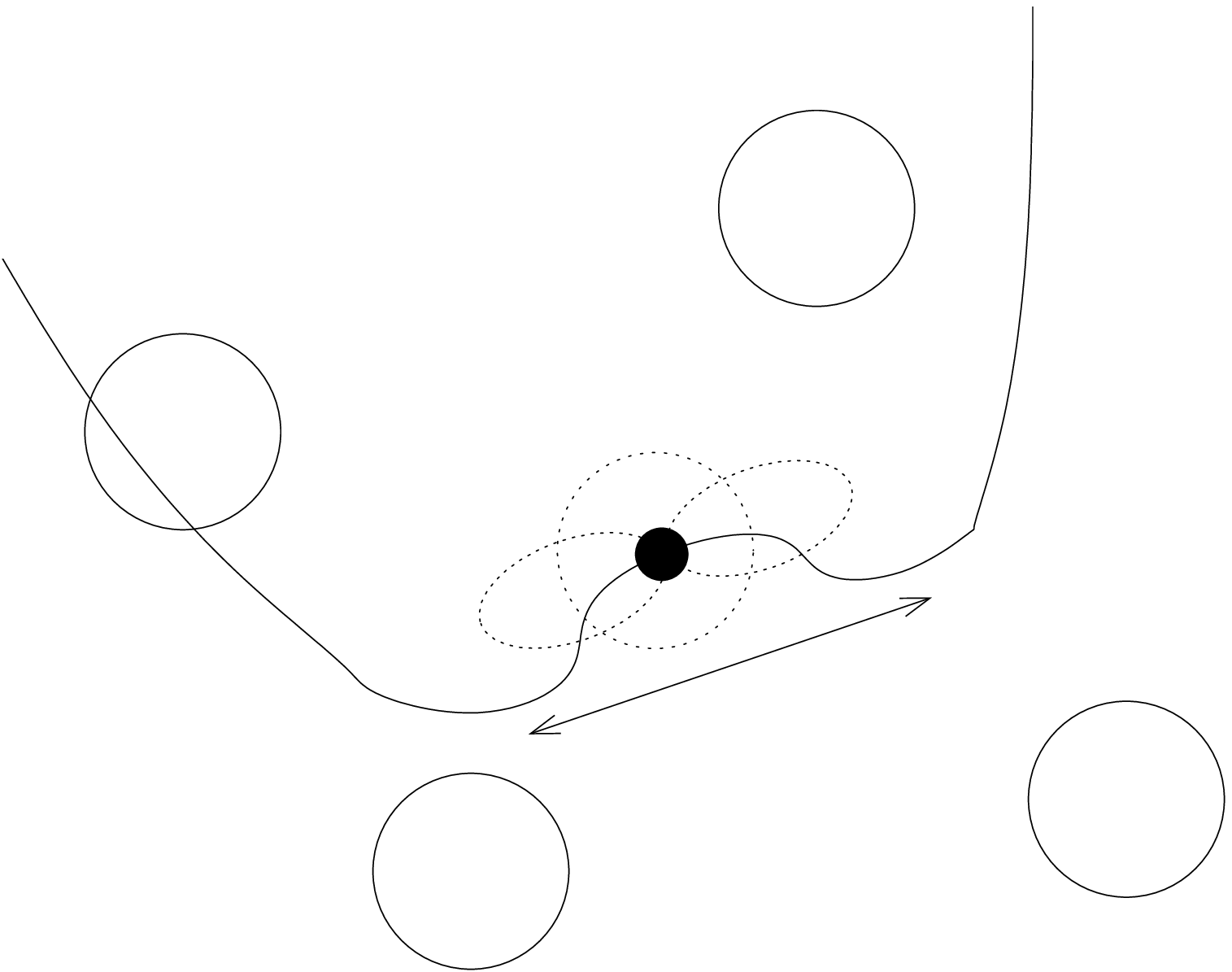}} \vskip 3mm
\centerline{\ \epsfysize 4cm \epsfbox{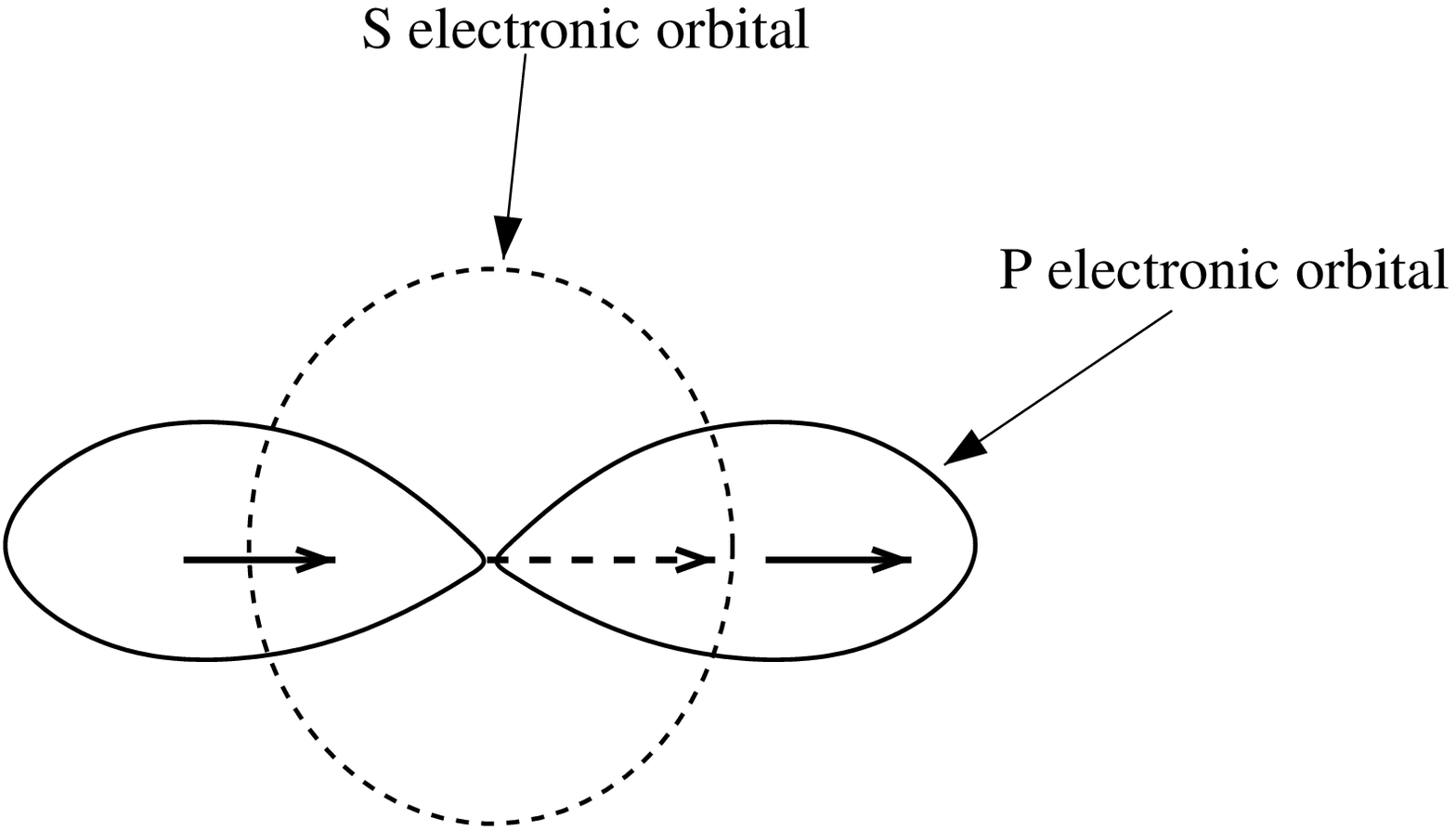}} \vskip 3mm
\caption{The highly directional atomic zero-point motion (solid
arrow) in the confining double-well potential of the surrounding
atoms (empty circles) in solid or liquid He. This motion of the
atom mixes the $s$ and $p$ level along the direction of motion.
Lower figure: The spins of the $s$ and excited $p$ electrons align
with the axis of the $p$-state, due to magnetic dipole interaction
of strength $\sim1.4$K.}
\end{figure}


\begin{references}
\bibitem{cold} Where the temperature is much smaller than the atomic zero-point
energy.
\end{references}
\end{document}